\documentclass{article}

\usepackage{arxiv}

\usepackage[utf8]{inputenc} 

\usepackage[english]{babel}
\usepackage{hyperref}       
\usepackage{url}            
\usepackage{booktabs}       
\usepackage{amsfonts}       
\usepackage{nicefrac}       
\usepackage{microtype}      
\usepackage{lipsum}		
\usepackage{graphicx}
\usepackage{natbib}
\usepackage{doi}
\usepackage{comment}
\usepackage{amsfonts}
\usepackage{amssymb}
\usepackage[table]{xcolor}
\usepackage{adjustbox}
\usepackage{float}
\usepackage{rotating}
\usepackage{array}
\usepackage{subcaption}
\usepackage{longtable}
\usepackage{enumerate}
\usepackage{enumitem}
\usepackage{multirow}
\usepackage{multicol}
\usepackage{fancyhdr} 
\usepackage{xcolor} 
\definecolor{cvteal}{HTML}{008080} 
\definecolor{cvblue}{HTML}{304263} 

\newcounter{rowno}
\addto{\captionsbrazil}{}

\hypersetup{
    colorlinks=false,
    hidelinks,
}

\usepackage{listings}
\definecolor{codegreen}{rgb}{0,0.6,0}
\definecolor{codegray}{rgb}{0.5,0.5,0.5}
\definecolor{codepurple}{rgb}{0.58,0,0.82}
\definecolor{backcolour}{rgb}{0.95,0.95,0.92}

\lstdefinestyle{mystyle}{
    backgroundcolor=\color{backcolour},   
    commentstyle=\color{codegreen},
    keywordstyle=\color{magenta},
    numberstyle=\tiny\color{codegray},
    stringstyle=\color{codepurple},
    basicstyle=\ttfamily\footnotesize,
    breakatwhitespace=false,         
    breaklines=true,                 
    captionpos=b,                    
    keepspaces=true,                 
    numbers=left,                    
    numbersep=5pt,                  
    showspaces=false,                
    showstringspaces=false,
    showtabs=false,                  
    tabsize=2
}
\title{Algorithmic Identity Based on Metaparameters: A Path to Reliability, Auditability, and Traceability}

\author{\href {https://orcid.org/0000-0001-9542-3732}{\includegraphics[scale=0.06]{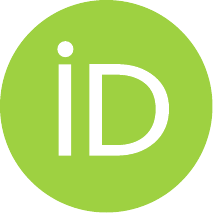}\hspace{1mm}Juliao Braga}\\
  Center for Mathematics, Computation and Cognition\\
  Federal University of ABC\\
  Santo André, SP, Brazil\\
  \texttt{juliao.braga@ufabc.edu.br}\\
  \And
  \href{https://orcid.org/0009-0009-7767-9625}{\includegraphics[scale=0.06]{orcid.pdf}\hspace{1mm}Percival Henriques}\\
  Brazilian Internet Steering Committee\\
  São Paulo, SP, Brazil\\
  \texttt{percival@cgi.br}\\
  \And
  \href{https://orcid.org/0000-0000-0000-0000}{\includegraphics[scale=0.06]{orcid.pdf}\hspace{1mm}Juliana C. Braga}\\
  Center for Mathematics, Computation and Cognition\\
  Federal University of ABC\\
  Santo André, SP, Brazil\\
  \texttt{juliana.braga@ufabc.edu.br}\\
  \And
  \href{https://orcid.org/0000-0000-0000-0000}{\includegraphics[scale=0.06]{orcid.pdf}\hspace{1mm}Itana Stiubiener}\\
  Center for Mathematics, Computation and Cognition\\
  Federal University of ABC\\
  Santo André, SP, Brazil\\
  \texttt{itana.stiubiener@ufabc.edu.br}\\
}

\begin{document}

\maketitle

\begin{abstract}
The use of algorithms is increasing across various fields such as healthcare, justice, finance, and education. This growth has significantly accelerated with the advent of Artificial Intelligence (AI) technologies based on Large Language Models (LLMs) since 2022. This expansion presents substantial challenges related to accountability, ethics, and transparency. This article explores the potential of the Digital Object Identifier (DOI) to identify algorithms, aiming to enhance accountability, transparency, and reliability in their development and application, particularly in AI agents and multimodal LLMs. The use of DOIs facilitates tracking the origin of algorithms, enables audits, prevents biases, promotes research reproducibility, and strengthens ethical considerations. The discussion addresses the challenges and solutions associated with maintaining algorithms identified by DOI, their application in API security, and the proposal of a cryptographic authentication protocol.

\vspace{0.5cm}
\noindent \textbf{Keywords:} DOI, AI Agents, Identification, Governance, Algorithms.
\end{abstract}

\section{Introduction}

The growing use of algorithms in various fields such as healthcare, justice, finance, and education brings significant challenges related to \textit{accountability}, ethics, and transparency. The increasing complexity and autonomy of algorithms, especially with the advancement of Artificial Intelligence (AI) \citep{russell:2021artificial} and Multimodal Large Language Models (MLLMs) \citep{iusztin2024llm}, demand effective mechanisms to ensure that these technologies are developed and used responsibly and ethically.

In this context, the identification of algorithms with the Digital Object Identifier (DOI) \citep{Liu:2021doi, shahane:2012digital} emerges as a promising tool to trace their origin, ensure reliability, and promote \textit{accountability} in cases of bias, errors, or harm caused by algorithms \citep{park2011examining}. This article explores the implications of identifying algorithms with DOI, addressing its benefits, challenges, and applications in different contexts, with particular focus on AI agents and MLLMs \citep{guo2024large, gates2025}.

The DOI is a unique and persistent identification system used to assign a code to digital objects. Created by the \textit{International DOI Foundation} (IDF), its original purpose was to guarantee citability in academic publishing. However, with the ubiquity of automated decision-making, the application of DOI to algorithm identification emerges as a necessary solution to improve governance.

\section{Taxonomy and Definition of the Identification Unit}

For the assignment of a DOI to be effective, it is imperative to precisely define the object being identified. The term "algorithm" encompasses everything from abstract logic to complex AI systems. We therefore propose a three-level taxonomy for identifier assignment:

\begin{enumerate}
    \item \textbf{Level 1 - Algorithmic Logic (Abstract Level):} Refers to the mathematical or logical conception of the solution (e.g., "Dijkstra's Algorithm") \citep{zermiani2022introduccao}. At this level, the DOI identifies the theoretical intellectual property and the authorship of the discovery, regardless of the programming language.
    \item \textbf{Level 2 - Reference Implementation (Code Level):} Refers to a specific implementation in source code (e.g., "Python Implementation v1.0 of Dijkstra"). Here, the DOI serves to ensure scientific reproducibility and code auditing, linking to versioned repositories.
    \item \textbf{Level 3 - AI Systems and Trained Models (Artifact Level):} In the context of \textit{Deep Learning} and LLMs, where logic is not explicitly programmed line by line but learned, the "algorithm" is insufficient as a unit of identification. In these cases, the DOI must be assigned to the \textbf{model artifact} (the file containing weights and parameters) together with its \textit{Model Card}.
\end{enumerate}

This distinction is crucial to address the challenge of opacity in AI \citep[p. 69]{henriques2025direito}. For AIs, DOI metadata must mandatorily reference the fingerprints (\textit{hashes}) of the training datasets and the ethical alignment reports \citep{sendin:1999funccoes}. The opacity of AI algorithms refers to the difficulty or impossibility of understanding how these systems reach their decisions. This occurs because many models operate as \textbf{black boxes}, hindering transparency, accountability, and social trust.

\section{Context and Comparative Analysis}

The proposal to use DOIs differs from already established technical mechanisms. Table \ref{tab:comparativo} presents a structural comparison demonstrating how the DOI fills the gap in institutional governance.

\begin{table}[!ht]
\centering
\caption{Comparison among identification mechanisms}
\begin{tabular}{|p{4.0cm}|p{1.7cm}|p{3.7cm}|p{4.8cm}|}
\hline
\multicolumn{1}{|c|}{\textbf{Mechanism}} & \multicolumn{1}{c|}{\textbf{Main Focus}} & \multicolumn{1}{c|}{\textbf{Advantage}} & \multicolumn{1}{c|}{\textbf{Limitation for Governance}} \\ \hline
 \textit{Git Hash} (SHA)  & Code Integrity & Ensures that the code has not been altered (mathematical immutability). & Does not provide human context, persistent authorship, or ethical metadata. It is volatile if the history is rewritten.  \\ \hline
 Digital Signature  & Author Authenticity & Ensures origin (who created it) via asymmetric cryptography.  & Does not offer academic citability, nor public access to documentation or purpose-related metadata. \\ \hline
 \textit{Software Heritage} (SWHID) & Long-term Archiving  & Universal preservation of source code over time. & Focuses on artifact preservation, not on institutional governance or explaining its social impact. \\ \hline
 Patents  & Legal Protection  & Intellectual property protection and monopoly of use.  & Slow, closed process focused on commercial exploitation, not on democratic or ethical transparency.  \\ \hline
 DOI Proposal & Governance and Citation & Rich metadata, persistent citability, and linkage with institutional responsibility. & Requires a central registration authority (RA) and maintenance costs. \\ \hline
\end{tabular}
\label{tab:comparativo}
\end{table}

Unlike digital signatures that prove \textit{who} signed, the proposed DOI acts as a \textbf{digital institution}, linking the technical object to a \textbf{responsibility charter}.

\section{Benefits of Using DOI}

Assigning DOIs to algorithms offers direct benefits for \textit{accountability} and transparency:
\begin{itemize}
    \item \textbf{Unique and Persistent Identification:} Facilitates citation in academic and technical publications.
    \item \textbf{\textit{Accountability} and Auditing:} Allows identification of the responsible institution in cases of bias or harm, facilitating legal audits.
    \item \textbf{Network Security:} In agent networks, ensures that only identified algorithms can interact.
\end{itemize}

\subsection{DOI in MLLMs and Traceability}
In the context of opaque MLLMs, the DOI functions as an institutional identity that enables:
\begin{itemize}[noitemsep]
    \item \textbf{Linking Hallucinations to the Source:} Trace exactly which version of the model generated problematic content.
    \item \textbf{Preventing “Model Washing”:} Verify whether a distributed model is genuine through the public key in its metadata.
    \item \textbf{Training Certification:} The DOI serves as a container to certify that the training data respected copyright.
\end{itemize}

\section{Autonomous Agents and Ethics}

In scenarios with non-deterministic agents \citep{wooldridge2009introduction}, the DOI becomes vital for tracking unpredictable actions. The ethical dimension aligns with what \citet{henriques2025direito} calls the \textbf{architecture of qualified transparency}, where transparency is not merely access to the code, but the understanding of its purpose, logic, and impact. The implementation of DOI operationalizes this framework by linking each identifier to structured explanations across these layers.

\section{Algorithms as Institutions}

If we consider algorithms as institutions \citep{almeida2024}, the DOI functions as an \textbf{algorithmic birth certificate}. This perspective finds support in digital constitutionalism \citep{henriques2025direito}, suggesting that the domestication of algorithmic power requires qualified transparency, where the DOI acts as a link between technical identity and the obligation of democratic explainability.

\section{APIs with DOI}

Each API should have a unique DOI, enabling usage tracking and version management. The \textit{handshake} between identified APIs would enhance security, ensuring that only trusted services exchange sensitive data. Furthermore, it would facilitate governance by regulatory bodies in critical sectors such as healthcare and finance.

\section{Technical Implementation: Metadata for Qualified Transparency}

The mere existence of a DOI does not guarantee trust; it is the richness of the metadata that matters. We propose an expanded mandatory metadata \textit{schema} for the registration of Algorithm-DOIs, going beyond the bibliographic standard:

\begin{itemize}[noitemsep]
    \item \textbf{Technical Identification Fields:}
    \begin{itemize}
        \item \texttt{Algorithm\_Type}: Deterministic, Stochastic, or Neural Network/AI.
        \item \texttt{Source\_Repo\_Hash}: Immutable link to the repository (e.g., SWHID).
        \item \texttt{Public\_Key}: The public key used for the authentication protocol.
    \end{itemize}
    
    \item \textbf{AI Governance Fields:}
    \begin{itemize}
        \item \texttt{Training\_Data\_Ref}: Reference to the datasets used (bias auditability).
        \item \texttt{Model\_Card\_URI}: Persistent link to the technical document on limitations and usage.
        \item \texttt{Ethical\_Alignment\_Report}: Statement of regulatory compliance.
    \end{itemize}
    
    \item \textbf{Social Transparency Fields:}
    \begin{itemize}
        \item \texttt{Citizen\_Explainability}: Non-technical summary of purpose and impact.
        \item \texttt{Responsible\_Institution}: Legally responsible entity.
    \end{itemize}
\end{itemize}

\section{Security and Authentication Protocol}

To ensure security in networks of autonomous agents, we propose a challenge-response protocol that leverages the DOI infrastructure:

\begin{enumerate}[label=DR\arabic*.,noitemsep]
    \item \textbf{Query:} The requesting agent (AA) retrieves the DOI metadata of the target agent (AB) and extracts its \texttt{Public\_Key}.
    \item \textbf{Challenge:} The AA sends a random \textit{token} to the AB.
    \item \textbf{Signature:} The AB signs the \textit{token} using its private key (never exposed) and returns the digital signature.
    \item \textbf{Validation:} The AA verifies the signature using the public key obtained from the DOI metadata.
\end{enumerate}

\begin{figure}[!ht]
\center
\includegraphics[scale=0.70]{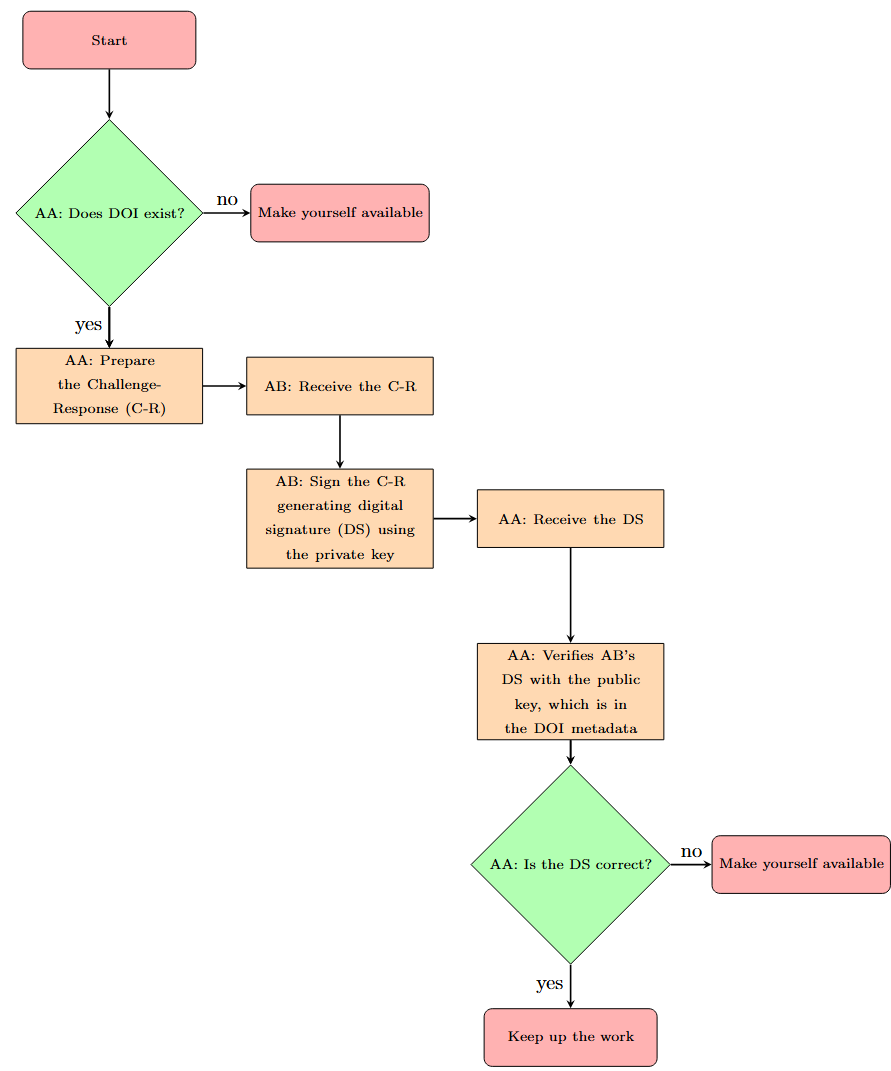}
\caption{The process of DOI verification and authentication of one algorithm by another} 
\label{fig:challenge-response-en}
\end{figure}

This process (Figure \ref{fig:challenge-response-en}), analogous to PGP \citep{zimmermann:1995}, ensures that only the entity holding the legal custody of the algorithm can operate under that DOI, preventing \textit{spoofing} of trusted models.

\section{Challenges of Maintenance and Versioning}

Software maintenance presents the challenge of the \textbf{Ship of Theseus} \citep{scatsas:1980theseus}: how much can an algorithm change before requiring a new DOI? We propose the following alternatives:

\begin{itemize}[noitemsep]
    \item \textbf{Incremental Changes:} Small fixes do not justify a new root DOI. The solution is semantic versioning in the metadata or suffixes (e.g., \texttt{10.xxx/algo.v1.2}).
    \item \textbf{Significant Changes:} Alterations in the core logic or complete retraining of an AI model require a new DOI, linked to the previous one through the metadata field \texttt{IsDerivationOf}.
\end{itemize}

\section{Limitations and Risks}

The implementation of a global system via DOI is not free of risks:
\begin{itemize}[noitemsep]
    \item \textbf{Costs and Centralization:} Unlike Git, DOI depends on Registration Agencies (RAs) and involves costs, which may create barriers for independent developers.
    \item \textbf{False Sense of Security:} DOI guarantees identity, not benevolence. It is essential to educate users that the identifier enables auditing, but it is not an automatic certificate of quality without verification of the metadata (\textit{Model Cards}).
\end{itemize}

\section{Future Work}

Future investigations should focus on the development of tools for automatic metadata generation from repositories (CI/CD), integration with platforms such as \textit{GitHub} and \textit{Hugging Face}, and the evaluation of Decentralized Identifiers (DIDs) \citep{Sabadello:22:DI} as a complementary alternative to DOI in order to reduce centralization.

\section{Conclusion}

The identification of algorithms with DOI opens up a range of possibilities to ensure accountability and transparency. By linking technical identity to governance metadata and a cryptographic authentication protocol, we transform the DOI from a mere catalog into an active tool for security and ethics. As emphasized by \citet{henriques2025direito}, the ultimate goal is not formal transparency, but qualified transparency that serves democratic values.

\setcitestyle{numbers}
\bibliographystyle{plainnat}
\bibliography{references} 

@ARTICLE{almeida2024,
  author = {Virgílio Almeida and Ricardo Fabrino Mendonça and Fernando Filgueiras},
  title = {{Thinking of Algorithms as Institutions: A route to democratic digital governance}},
  journal = {Communications of the ACM},
  year = {2024},
  volume = {67},
  number = {1},
  pages = {20-23},
  doi = {10.1145/3680411},
}

@misc{gates2025,
  author       = {Bill Gates},
  title        = {{AI Agents}},
  howpublished = {\url{https://www.gatesnotes.com/meet-bill/tech-thinking/reader/ai-agents}},
  year         = {2025},
  note         = {Accessed: January 11, 2025}
}

@article{guo2024large,
  title={{Large language model based multi-agents: A survey of progress and challenges}},
  author={Guo, Taicheng and Chen, Xiuying and Wang, Yaqi and Chang, Ruidi and Pei, Shichao and Chawla, Nitesh V and Wiest, Olaf and Zhang, Xiangliang},
  journal={arXiv preprint arXiv:2402.01680},
  year={2024}
}

@book{henriques2025direito,
  author    = {Henriques, Percival},
  title     = {{Direito à Realidade: Por um Constitucionalismo Digital para o Brasil}},
  publisher = {Editora Publius},
  address   = {Recife},
  year      = {2025},
  isbn      = {978-65-85007-50-4}
}

@book{iusztin2024llm,
  title={{LLM Engineer's Handbook: Master the art of engineering large language models from concept to production}},
  author={Iusztin, Paul and Labonne, Maxime},
  year={2024},
  month={out},
isbn={978-1-83620-007-9},
  publisher={Packt Publishing Ltd.}
}

@article{Liu:2021doi,
title = {{Digital Object Identifier (DOI) and DOI Services: An Overview}},
author = {Jia Liu},
pages = {349--360},
volume = {71},
number = {4},
journal = {Libri},
doi = {10.1515/libri-2020-0018},
year = {2021},
lastchecked = {2025-12-13}
}

@article{park2011examining,
  title={{Examining success factors in the adoption of digital object identifier systems}},
  author={Park, Sungbum and Zo, Hangjung and Ciganek, Andrew P and Lim, Gyoo Gun},
  journal={Electronic commerce research and applications},
  volume={10},
  number={6},
  pages={626--636},
  year={2011},
doi={https://doi.org/10.1016/j.elerap.2011.05.004},
  publisher={Elsevier}
}

@book{russell:2021artificial,
  title={{Artificial Intelligence: A Modern Approach}},
  author={Russell, Stuart and Norvig, Peter},
edition={4a.},
  year={2021},
  publisher={Pearson},
}

@TechReport{Sabadello:22:DI,
  author      = "Markus Sabadello and Drummond Reed and Manu Sporny and Amy Guy",
  title       = {{Decentralized Identifiers ({DIDs}) v1.0}},
  month       = jul,
  note        = "https://www.w3.org/TR/2022/REC-did-core-20220719/",
  year        = "2022",
  bibsource   = "https://w2.syronex.com/jmr/w3c-biblio",
  type        = "{W3C} Recommendation",
  institution = "W3C",
}

@article{scatsas:1980theseus,
 ISSN = {00032638, 14678284},
 author = {Theodore Scaltsas},
 journal = {Analysis},
 number = {3},
 pages = {152--157},
 publisher = {[Analysis Committee, Oxford University Press]},
 title = {{The Ship of Theseus}},
 urldate = {2025-12-13},
 volume = {40},
 year = {1980},
doi = {10.2307/3327668},
}

@mastersthesis{sendin:1999funccoes,
  title={{Fun{\c{c}}{\~o}es de Hashing Criptogr{\'a}ficas}},
  author={Sendin, Ivan da Silva},
  year={1999},
  school={Unicamp, SP. BR}
}

@article{shahane:2012digital,
  title={{Digital Object Identification: An Overview}},
  author={Shahane, Shraddha and Chauhan, Manjula},
  journal={Information Technology Applications: Strategies, Issues and Challenges},
  pages={140},
year={2012},
doi={10.13140/RG.2.2.18564.58242}
}

@book{wooldridge2009introduction,
  title={{An introduction to multiagent systems}},
  author={Wooldridge, Michael},
  year={2009},
  publisher={John Wiley \& Sons},
}

@article{zermiani2022introduccao,
  title={{Uma Introdu{\c{c}}{\~a}o ao Algoritmo de Dijkstra}},
  author={Zermiani, Aline Thaise and Fidalgo, Felipe Delfini Caetano},
  journal={Proceeding Series of the Brazilian Society of Computational and Applied Mathematics},
  volume={9},
  number={1},
  year={2022},
doi     = {10.5540/03.2022.009.01.0007}
}

@book{zimmermann:1995,
author = {Zimmermann, Philip R.},
title = {{The official PGP user's guide}},
year = {1995},
isbn = {0262740176},
publisher = {MIT Press},
address = {Cambridge, MA, USA}
}
\end{document}